\documentclass[suppldata]{interact}
\pdfoutput=1
\usepackage[english]{babel}
\usepackage[numbers]{natbib} 
\usepackage{natbib}
\usepackage{graphicx}
\usepackage{hyperref}
\usepackage{multirow}
\usepackage{booktabs} % For prettier tables
\usepackage{pifont} % For tick marks
\newcommand{\cmark}{\ding{51}} % Command to insert a tick mark
\newcommand{\MIV}{MIV }
\usepackage{makecell}
\usepackage{subcaption}
\usepackage[T1]{fontenc}
\usepackage{lmodern}

\usepackage{geometry}

\usepackage{enumitem}

\usepackage{tikz}
\usepackage{tcolorbox}
\usetikzlibrary{shapes.geometric, arrows, positioning}

\usepackage{listings} % For code listing
\usepackage{xcolor} % Required for listings color

\lstdefinestyle{mystyle}{
    basicstyle=%
    \ttfamily
    \color{blue}%
    \lst@ifdisplaystyle\large\fi,
    backgroundcolor=\color{white},   
    commentstyle=\color{teal},
    keywordstyle=\color{magenta},
    numberstyle=\tiny\color{gray},
    stringstyle=\color{purple},
    basicstyle=\ttfamily\footnotesize,
    breakatwhitespace=false,         
    breaklines=true,                 
    captionpos=b,                    
    keepspaces=true,                 
    numbers=left,                    
    numbersep=5pt,                  
    showspaces=false,                
    showstringspaces=false,
    showtabs=false,                  
    tabsize=2,
    frame=single
}

\lstdefinelanguage{JSON}{
    string=[s]{"}{"},
    stringstyle=\color{purple},
    comment=[l]{:},
    commentstyle=\color{black},
    numbers=left,
    numberstyle=\small,
    stepnumber=1,
    numbersep=8pt,
    showstringspaces=false,
    breaklines=true,
    frame=lines,
    basicstyle=\ttfamily\footnotesize,
    frame=single
}

\usepackage{parskip}  % Removes paragraph indentation
\newcommand{\mypar}[1]{%
  \par\noindent\textbf{#1}\space
}

\makeatother

\title{Model Input Verification of Large Scale Simulations}

\author{
\name{Rumyana Neykova %\textsuperscript{a}\thanks{CONTACT Rumyana Neykova.
and Derek Groen\textsuperscript{a}}
\affil{\textsuperscript{a}Brunel University London}
}

\begin{document}
\maketitle

\begin{abstract}
\begin{comment}
The importance of simulation modeling lies in its ability to predict and analyze complex systems, making it crucial to verify the correctness of these simulations. One often overlooked aspect of this process is model input verification (MIV), ensuring that the input data adheres to specified constraints and accurately represents the real-world scenarios being modeled. 
%It is a step that is as foundational as it is critical, directly impacting the integrity of the simulations. 
%Mismanagement of this phase can lead to erroneous outcomes and research setbacks.  
Incorrect input files can introduce various issues, such as invalid or out-of-range values, missing data, and format inconsistencies.
%, all of which can significantly distort simulation results and undermine the validity of the conclusions. 
These errors can have cascading effects, ranging from crashing the simulation to going unnoticed and distorting the final results, hence undermining the validity of the conclusions. 
\end{comment}
Reliable simulations are critical for analyzing and understanding complex systems, but their accuracy depends on correct input data.
%Accurate simulations are essential for analyzing complex systems, yet their reliability heavily depends on correct input data. 
Incorrect inputs such as invalid or out-of-range values, missing data, and format inconsistencies can cause simulation crashes or unnoticed result distortions, ultimately undermining the validity of the conclusions.
%Model input verification (MIV) is a critical but often overlooked process that ensures input data validity. 
\setlength{\parskip}{0pt}
 This paper presents a methodology for verifying the validity of input data in simulations, a process we term model input verification (MIV). We implement this approach in FabGuard, a toolset that uses  established data schema and validation tools for the specific needs of simulation modeling. We introduce a formalism for categorizing MIV patterns and offer a streamlined verification pipeline that integrates into existing simulation workflows. FabGuard's applicability is demonstrated across three diverse domains: conflict-driven migration, disaster evacuation, and disease spread models. We also explore the use of Large Language Models (LLMs) for automating constraint generation and inference. In a case study with a migration simulation, LLMs not only correctly inferred 22 out of 23 developer-defined constraints, but also identified errors in existing constraints and proposed new, valid constraints. Our evaluation demonstrates that MIV is feasible on large datasets, with FabGuard efficiently processing 12,000 input files in 140 seconds and maintaining consistent performance across varying file sizes.

\end{abstract}
\section{Introduction}
Simulations have become an indispensable tool across various scientific disciplines, offering insights into complex systems ranging from epidemiology and environmental science to social dynamics and engineering in many different ways~\cite{epstein2008}. Recent advancements in computational power and data analytics have enabled researchers to develop and apply more realistic and actionable simulation approaches, and deliver benefits in a growing number of areas. For instance, in epidemiology, simulations have been pivotal in modeling the spread of infectious diseases like COVID-19 \cite{ferguson2020,mahmood2022}, while in environmental science, they have provided insights into ecosystem interactions and biodiversity under changing climate conditions \cite{geary2020,dada2011,jahani2023}. 
%The growing complexity and accessibility of simulation tools have not only expanded their application across traditional fields but have also made them instrumental in addressing some of the most pressing challenges of our time.

As these models increasingly influence critical decision-making processes, ensuring their reliability and reproducibility has become paramount \cite{coveney_big_2016}. The importance of reproducibility in simulation modeling is supported by the growing emphasis on open-source practices in scientific computing in general. The open-source software movement has played a crucial role in promoting software sustainability and reproducibility, particularly in scientific simulations \cite{coveney_reliability_2021,coveney_reliability_2021}. Free and Open Source Software (FOSS) enhances the longevity and adaptability of software projects, ensuring that simulation tools remain accessible and maintainable over time \cite{coveney2021}. 
%This trend towards openness and transparency in scientific software development aligns with the broader push for reproducible research in computational science \cite{y}.
Initiatives such as the Journal of Open Source Software (JOSS) \cite{smith_journal_2018} and the increasing number of journals requiring code availability demonstrate the scientific community's recognition of the critical role that software plays in research reproducibility. In the context of simulation modeling, open-source practices not only facilitate peer review of the underlying code but also enable researchers to verify and build upon existing models, fostering cumulative scientific progress \cite{benureau_re-run_2018}. However, despite the abovementioned advancements, significant challenges remain. One of the primary barriers to reproducible research is that many of the tools required for reproducibility, such as version control, unit testing, and automation, are often seen as being of interest only to professional coders \cite{alhozaimy2017towards}. This perception gap highlights the need for solutions that can make these essential practices more accessible and relevant to domain experts who may not have extensive software engineering backgrounds.

While verification, validation and uncertainty quantification has received clear attention from researchers in recent years (see e.g. Coveney et al.~\cite{coveney2021}), a crucial and often overlooked aspect of ensuring simulation reliability and reproducibility is the process of validating and verifying simulation model input data. In particular, few generic approaches exist that verify that model input data adheres to specified constraints that ensure correct simulation execution and that it correctly represents the real-world scenarios being modeled. We call this process Model Input Verification (MIV). This step is important in guarding against simulation results being corrupted by human data input errors or poorly formatted raw input, as well as aiding in the prevention of cascading errors or crashes that can arise from such flawed or misrepresented inputs. The implications of inadequate input verification in simulation modeling can be severe. For instance, in 1999 a mistaken unit type in one of the ground software submodels led to the NASA Mars Climate Orbiter having an incorrect trajectory and burning up in the Martian atmosphere~\cite{stephenson1999}. Similarly, in Flee migration simulations it occasionally happens that developers retrieve GPS coordinates for locations in their simulation, and accidentally insert the coordinates of identically named places that reside in an entirely different country.

Recent years have seen a growing emphasis on testing data and ensuring data quality, forming the basis for test-driven data analysis and 'unit tests' for data \cite{10.14778/3229863.3229867}. Libraries such as Pandera \footnote{https://www.union.ai/pandera}, Great Expectations \footnote{https://greatexpectations.io/}, and Cerberus \footnote{https://pypi.org/project/Cerberus/} have emerged to verify data constraints and validate schemas. These tools have proven valuable in fields like data science and business intelligence, where they help maintain data integrity and catch errors early in the analysis pipeline \cite{bantilan_pandera_2020}.

However, the development of simulation models often occurs in environments quite different from traditional software engineering. Typically, these simulations are created by domain experts - scientists, researchers, and analysts - who, while highly skilled in their fields, may not have extensive programming backgrounds \cite{merali_computational_2010}. This gap between domain expertise and software engineering practices has long been a challenge in ensuring the reliability and verifiability of scientific simulations \cite{bhthacker_concepts_2004, roy_comprehensive_2011}. Moreover, the tools and approaches for data validation have not been widely translated to simulation modeling and are often unavailable to simulation practitioners. Simulation inputs require constraints that go beyond simple data validation. For example, in agent-based models of population displacement, input verification must ensure not only that population values are non-negative but also that the sum of populations across different locations matches the total simulated population. Additionally, temporal consistency in input data is crucial; in disease spread models, the order and timing of intervention measures must align with the simulation timeline.

To address the aforementioned challenges and improve the reliability of simulation models, this paper presents FabGuard\footnote{Upon acceptance, the code will be made available on zenodo}, integrated set of tools and methods
for Model Input Verification. Our work is guided by several key research questions: How can we effectively adapt existing data validation tools to the unique needs of simulation modeling? What are the types of input verification constrained that a model should support?
How can we incorporate input verification into existing simulation workflows? 
To what extent can Large Language Models (LLMs) assist in inferring and generating input verification rules to help with adoption of MIV tools? 
%How can we develop a verification framework that handles complex, multi-dimensional constraints typical in simulations? 

In addressing these questions, our work offers several novel contributions to the field. 
%We have formalized the main constraints and developed a framework for identifying input verification requirements, providing a structured approach to this aspect of simulation modeling. This formalism presents a set of constraint types across various dimensions of simulation input data, offering a systematic way to address verification needs. Building on this foundation, we introduce a streamlined verification pipeline that can be easily integrated into the Continuous Integration/Continuous Deployment (CI/CD) workflows of simulation models, promoting automated input verification. To demonstrate the practical applicability of our approach, we showcase the use of FabGuard across three diverse simulation domains: conflict-driven migration, disaster evacuation, and disease spread models. This application highlights the adaptability of off-the-shelf tools for input verification in complex simulation scenarios. Finally, we present the first study examining the suitability of LLMs for constraint generation and inference in the context of Model Input Verification, demonstrating their potential to lower the learning curve for simulation practitioners. 

\begin{itemize}[noitemsep, topsep=0pt]
\item[\S~\ref{sec:workflow}] \textbf{Introduces Fabguard, a streamlined verification pipeline} that can be easily integrated into CI/CD workflows of simulation models, promoting automated input verification.
\item[\S~\ref{sec:formalism}] \textbf{Formalizes model input verification requirements} for simulation modeling. We present a framework categorizing constraint types across various dimensions of simulation input data, offering a systematic approach to address verification needs.
\item[\S~\ref{sec:exemplars}] \textbf{Demonstrates the practical applicability of FabGuard} across three diverse simulation domains: conflict-driven migration, disaster evacuation, and disease spread models. This showcases the adaptability of off-the-shelf tools for input verification in complex simulation scenarios.
\item[\S~\ref{sec:llm}] \textbf{Presents the first study examining the suitability of LLMs} for constraint generation and inference in the context of Model Input Verification, demonstrating their potential to lower the learning curve for simulation practitioners.
\item[\S~\ref{sec:evaluation}] \textbf{Evaluates FabGuard's performance} providing insights into its scalability and efficiency in various scenarios.
\end{itemize}

%The rest of this paper is organized as follows: Section~\ref{sec:overview} provides an overview of Model Input Verification, introduces FabGuard, and presents a formalism for categorizing MIV patterns. Section~\ref{sec:exemplars} presents exemplars demonstrating FabGuard's application in various simulation contexts. Section~\ref{sec:llm} explores the use of Large Language Models for constraint inference and generation. Section~\ref{sec:evaluation} evaluates FabGuard's performance through microbenchmarks and a real-world case study. 
Section~\ref{sec:related} discusses related work, and Section~\ref{sec:conclusion} concludes with a summary of contributions and future directions.
\section{Related Work}
\label{sec:related}
The problem of reproducibility in computational science has been identified as a critical issue \cite{coveney_big_2016}, and there are ongoing efforts to address it \cite{coveney_reliability_2021}. 
Automated testing is needed to systematically verify computer simulations, a precondition to ensuring that the results they produce are sufficiently robust to inform decision-making in the real world~\cite{coveney_when_2021}.
This section contextualizes the role of input verification within the broader domain of simulation modeling and further explores solutions in the fields of data analytics and data workflows, which face similar challenges.
{\bf Verification of simulations} is crucial for ensuring that computational models accurately represent real-world scenarios and for enhancing reproducibility. Various approaches and tools have been developed to enhance this process. Code verification focuses on identifying programming errors and verifying numerical algorithms through Software Quality Assurance (SQA) procedures, ensuring software reliability and consistency \cite{bhthacker_concepts_2004}. Comprehensive frameworks for Verification, Validation, and Uncertainty Quantification (VVUQ) further improve predictive capabilities by incorporating methods to estimate and propagate uncertainties through models \cite{roy_comprehensive_2011}.
Several frameworks and large toolkits have been developed to address these challenges. For example, the VECMA toolkit \cite{vecmatk} offers a suite of tools for verification, validation, sensitivity analysis, and uncertainty quantification. Within VECMAtk, EasyVVUQ \cite{easyVVUQ} streamlines VVUQ for computationally expensive simulations and extensive sampling spaces. FabSim3 \cite{fabsim3}, a Python-based automation toolkit, reduces human effort in simulation-based research and provides an auto-validation tool for comparing simulation accuracy. The Model Verification Tools (MVT) framework \cite{MVT} offers mechanisms for VVUQ assessment of agent-based models, including sensitivity analysis techniques. Uncertainpy \cite{tennoe_uncertainpy_2018} facilitates robust simulation modeling by offering uncertainty quantification and sensitivity analysis using quasi-Monte Carlo and polynomial chaos expansions methods. For a comprehensive overview of many works on validation and verification, especially for uncertainty quantification, readers are directed to \cite{coveney_reliability_2021}.

Beyond these specific tools, there are more general works addressing various aspects of simulation verification and validation. Gundersen \cite{gundersen_fundamental_2021} emphasize the importance of transparency and openness as key drivers for reproducibility. \cite{roungas2018} address the challenge of selecting appropriate V\&V methods due to the abundance of available techniques, proposing a methodology for choosing the most suitable methods based on simulation characteristics. In the realm of high-performance computing, Encinas et al. \cite{encinas2019} present a simulation model of HPC I/O systems using Agent-Based Modeling and Simulation (ABMS), providing insights into I/O performance behavior in different configurations. Farrell et al. \cite{farrell2011} highlight the importance of automated continuous testing in numerical modeling, demonstrating significant improvements in code quality and 
programmer efficiency. \cite{sinisi2021} address interoperability challenges in Cyber-Physical System (CPS) simulation, presenting an implementation of FMI 2.0 functions for improving efficiency in simulation-based V\&V. These diverse approaches collectively contribute to ongoing efforts to improve the reliability, efficiency, and reproducibility of simulation-based research across various domains.

%Although there are many initiatives to incorporate VVUQ,  
%Code solution is mentioned as instrumental in the verification process, 
%none address the 
%many of the state-of-the-art tools for simulation modelling focus primarily on UQ. Neither of the above mentioned tools address input verification. 

Despite these advancements, there remains a notable gap in addressing model input verification. Most existing tools and frameworks focus on verifying simulation code, quantifying uncertainties, or validating outputs, rather than verifying input data. 
%For instance, EasyVVUQ, while comprehensive in uncertainty quantification, does not perform model input verification \cite{easyVVUQ}. This gap is critical, as input errors can propagate through the entire simulation process, potentially leading to unreliable results. 
The current paper addresses this crucial aspect of simulation reliability by focusing specifically on model input verification, thus complementing existing VVUQ approaches.

\textbf{Data validation and verification} has gained significant attention in data science and machine learning communities. Schelter et al. \cite{Shelter} introduced the concept of "unit tests" for data, providing a framework for describing data constraints. This has spurred research into data schema generation, inference, and validation techniques for complex machine learning applications \cite{10.14778/3137765.3137789, Schelter2018, 47352}. Modern machine learning platforms now incorporate explicit data validation components, addressing issues such as data drift, model performance degradation, and input data quality \cite{jha_data_2019, smith_journal_2018, shankar_automatic_2023, wong_mlguard_2023, siddiqi_saga_2023, patel_data-centric_2023}.

The growing emphasis on data quality and schema verification has led to the development of several tools and libraries aimed at streamlining these processes. Great Expectations \cite{great_expectations} has emerged as a popular tool for data validation and documentation, allowing users to express their data expectations in a declarative manner and facilitating automated testing of data quality. Pandera \cite{bantilan_pandera_2020} provides a flexible and expressive API for performing data validation on pandas DataFrames, enabling the definition of schemas with column-level and dataframe-level validation rules, including complex statistical checks.
Other tools like Cerberus \cite{cerberus} offer similar functionality, reflecting a broader trend towards more robust, automated approaches to data validation across various domains. The TDDA Python module \footnote{https://github.com/tdda/tdda} supports test-driven data analysis through various tools, including Reference Testing for managing complex data analysis pipeline tests and tools for discovering, validating, and detecting anomalies in data constraints.

These developments in data validation techniques and tools provide a strong foundation for addressing similar challenges in the simulation domain. While the focus of these works has primarily been on data science and machine learning applications, many of these approaches and tools can be adapted or repurposed for simulation input verification.
%The development of specialized tools and frameworks, as discussed in works on test-driven data analysis, points to a future where input verification is a standard and automated part of the simulation modeling process. 
In the context of the extensive literature on VVUQ for simulation models, input verification is acknowledged but still not deeply explored. However, as simulations become more complex and as reproducibility becomes a more pressing concern in scientific research, the role of input verification will become ever more prominent.  
%gaining recognition. 
%\todo{Rumi: Add works... from other domains different that simulation modelling}

%In conclusion, input verification is an essential yet often understated element of the VV framework in simulation modeling. It is a critical first line of defense against errors that can undermine the validity of a simulation, and its importance is increasingly recognized in both practical applications and academic discussions within the field.

\section{MIV Overview}
\label{sec:overview}
This section provides an overview of Model Input Verification (MIV), its importance in simulation modeling, and introduces FabGuard as a comprehensive toolset for implementing MIV. We begin by explaining the concept and significance of MIV. We then present a formalism for categorizing different types of input verification tasks, which serves as a framework for understanding and implementing MIV processes. Finally, we introduce FabGuard, detailing its architecture and key features.

Model Input Verification is an important step in the simulation modeling process, ensuring that input data adheres to specified constraints and accurately represents the real-world scenarios being modeled. In essence, MIV allows users to write tests that check whether input files meet specific requirements and satisfy a set of predefined constraints. These tests help prevent cascading errors that can arise from flawed or misrepresented inputs, enhancing the reliability and reproducibility of simulation results. Common MIV tasks include checking data types, value ranges, inter-column relationships, and cross-file consistency.

\begin{figure}
    \centering
    \includegraphics[scale=0.5]{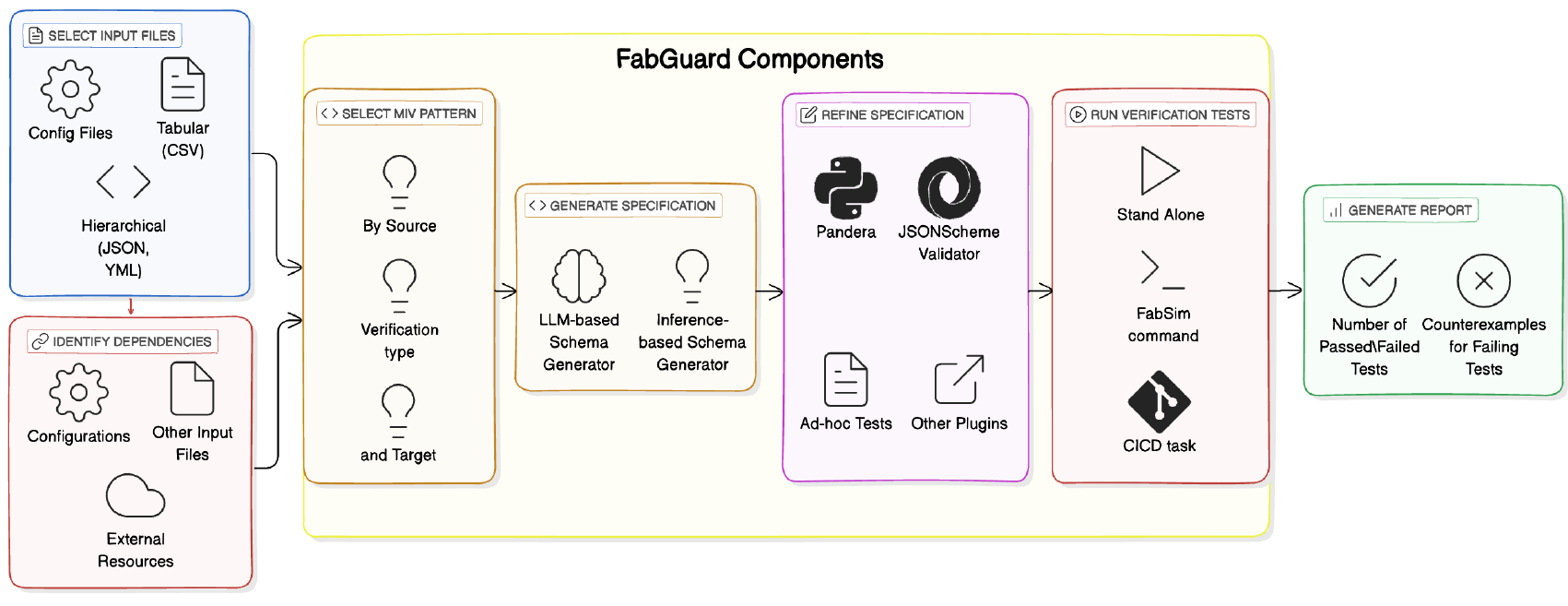}
    \caption{Methodology for model input files verification}
    \label{fig:methodology}
\end{figure}

To illustrate the toolchain  and the main ideas behind model input verification, we use as a running example an agent-based simulation, called Flee ~\cite{suleimenova2017},  designed for modeling displacement and migration patterns. Flee enables researchers to create simulations based on conflict and disaster scenarios, helping to predict how populations move in response to various crises. It has been utilized in major research initiatives such as the EU-funded HiDALGO and ITFLOWS projects. Within Flee, agents move across a location graph defined by two primary input CSV files: locations.csv, which defines the nodes of the graph representing various locations such as towns, camps, and conflict zones, and routes.csv, which defines the edges of the graph, representing possible paths between locations.

%Additional input files include closures.csv, which specifies border closures or route disruptions during the simulation, and simsetting.yml, which configures the set of assumptions used in the simulation, introduced in Flee version 3 \cite{flee}. Errors in these input files can lead to various issues, including inaccuracies in the simulation results, agents getting stuck in certain locations, or Flee crashing altogether.

\subsection{\MIV Workflow}
\label{sec:workflow}

Fig.~\ref{fig:methodology} depicts the high-level methodology of writing MIV tests. Notably, the first two stages - selecting input files and identifying dependencies - are manual processes performed by the user. These initial steps are important for establishing the context and scope of the verification process. FabGuard is designed to support and automate the subsequent stages, providing a plugin-based architecture that accommodates various input file formats and validation methods.

%The methodology of FabGuard is structured into several key stages to streamline the verification process, as seen in Fig.~\ref{fig:methodology}.

\textbf{Selection of Input Files:} In the initial stage the user should select input files to verify. The format and content will vary and are simulation-specific. The files are categorized into configuration files, which provide necessary settings for running simulations, and input files that supply the data required to execute processes.  For instance, in the Flee simulation tool, the input files might include \texttt{locations.csv} and \texttt{routes.csv} which contain tabular data, while the configuration file is \texttt{simsettings.yaml} and contains key-value pairs of simulation parameters. 
%The \texttt{locations.csv} file contains details such as town names and coordinates, while the \texttt{routes.csv} file describes the connectivity between these towns.

\textbf{Identifying Dependencies:} In the next stage, the user must identify dependencies essential for parameterizing the inputs. This involves configurations that require specific settings, supplementary input files providing context, and external resources such as databases or APIs needed for validation. For example, if \texttt{simsetting.yml} sets the simulation to start on January 1, 2023, any closure events in \texttt{closures.csv} with earlier dates should be flagged as invalid.

%In Flee, the \texttt{closures.csv} file might require validation against real-world border closure events, . %accessed through an API to ensure simulations reflect actual scenarios.

\textbf{Generating Specifications:} 
%FabGuard then generates specifications that act as rules or schemas for validating input files. This is achieved through various methods, including the LLM-based schema generator, which automatically generates schemas based on input data, aiding in understanding complex file structures. An inference-based schema generator infers schemas by analyzing file content, ensuring that data like \texttt{locations.csv} meets structural requirements. Manual input also allows users to create or refine schemas, ensuring they match specific simulation needs.
Once the user has identified the input files for verification and their potential dependencies, they can begin writing input verification tests. FabGuard supports two off-the-shelf libraries for schema validation, depending on the type and format of the data -- \texttt{Pandera} and \texttt{jsonschema}. The former is a  library for defining schemas and validating pandas DataFrames; which allows users to define column-level and dataframe-level validation rules, including data types, value ranges, and custom checks. The latter is a lightweight way to test your YAML/JSON files based on how they conform to a defined schema.
FabGuard provides a thin wrapper over both \texttt{Pandera} and \texttt{jsonschema} libraries, enabling integration with simulation tools, LLMs, and providing consistent documentation. Users can start writing tests using the library that best suits their case.

However, writing these tests can be a tedious process that requires programming skills, potentially hindering the tool's applicability. To address this, we have explored two potential ways to bootstrap this stage:
\begin{enumerate}
\item{\textbf{Schema Generators:}} FabGuard supports built-in schema generators - a custom yaml schema generator, and a Pandera inference module. These tools can automatically infer basic constraints such as data types, minimum and maximum values for most files. While not comprehensive, they create useful scaffolding that can later be refined by users. For instance, a schema generator might infer that the 'population' column in \texttt{locations.csv} should contain non-negative integers.
%\textbf{TODO: Explain that there are many such tools, give citation, they can be integrated in the tool, or used independently}

\item \textbf{Large Language Models (LLMs):} As reported in Section \ref{sec:llm}, we have explored the use of LLMs for constraint generation and inference. Our findings indicate that LLMs can not only create the scaffolding of the main tests but also suggest and infer novel constraints. For example, an LLM might suggest that the sum of populations across all locations should match the total simulation population, a constraint that might not be immediately obvious to users.
\end{enumerate}

These automated approaches serve as a starting point, providing a basic scaffolding which can then be refined and expanded by domain experts. This stage significantly lowers the barrier to entry for using FabGuard, making it more accessible to researchers who may not have extensive programming experience.
%By combining off-the-shelf libraries, custom wrappers, schema generators, and LLM-based inference, FabGuard offers a framework for generating input verification specifications. 

\textbf{Refining Specifications:} 
The test should be further refined, and most importantly, verified. This stage is important, especially if automated inference tools were used in the previous steps. As outlined in Section~\ref{sec:llm}, some constraints, although they can be inferred, may require adjustments to accurately reflect the simulation's requirements. For example, in Flee, an inferred constraint might correctly identify that the 'population' field should be non-negative, but may need refinement to specify that conflict zones must have a non-zero population while other location types can have zero population. 

It's important to note that the previous stages of automated inference are optional. Developers can choose to write all tests from scratch, tailoring them precisely to their simulation's needs. Additionally, custom checks can be written for specific validation scenarios not covered by standard tools or inferred constraints. For instance, in Flee, a custom check might be needed to ensure that all routes listed in routes.csv correspond to actual connections between nodes specified in locations.csv, a relationship that may not be captured by automated inference tools.

\textbf{Running Tests:} FabGuard supports several ways for running the input verification tests. It is currently integrated with FabSim3 \cite{fabsim3}, a Python-based automation toolkit for scientific simulation and data processing workflows. This integration allows users to run FabGuard tests as part of simulation workflows within FabSim3 or execute them independently for focused input verification. Furthermore, FabGuard tests can be incorporated into Continuous Integration/Continuous Deployment (CI/CD) pipelines, such as GitHub Actions, enabling ongoing automated validation. 

%For example, ensuring that \texttt{closures.csv} accurately reflects temporary route closures prevents agents from accessing closed paths during the simulation.

\textbf{Report Generation:} In the final stage, FabGuard generates a report detailing test outcomes, including the number of passed and failed tests. Counterexamples for failing tests are provided, highlighting where and why certain tests failed and offering insights for corrective actions. If \texttt{locations.csv} fails validation due to missing entries, the report pinpoints these omissions, as well as the the exact rows and values which do not satisfy the constraints.

%While this workflow provides a practical approach to implementing MIV, it's essential to understand the underlying conceptual framework that guides these verification processes. To this end, we have developed a formalism that categorizes different types of verification tasks, providing a structured way to think about and implement MIV across various simulation domains.

%This systematic approach ensures thorough validation of input files, accurate identification of dependencies, and comprehensive specification generation. By integrating with various testing and CI/CD tools, FabGuard provides a framework for ensuring the integrity and correctness of input files, thereby enhancing the reliability of simulations and processes that depend on these files. Thi

\subsection{\MIV Conceptual Overview}
\label{sec:formalism}

\begin{figure}
    \centering
    \includegraphics[scale=0.5]{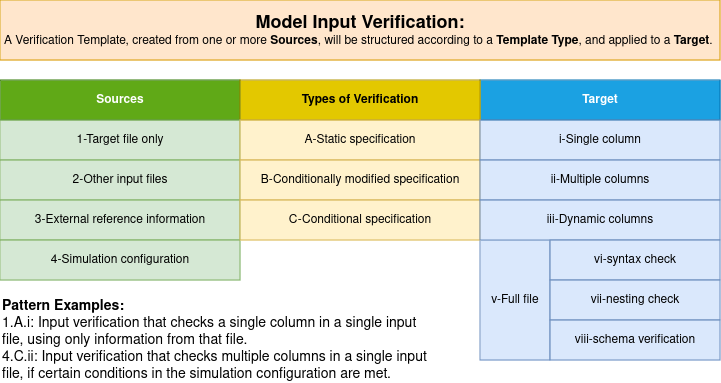}
    \caption{Overview of the MIV formalism}
    \label{fig:formalism}
\end{figure}

The MIV workflow described above encompasses a wide range of verification tasks, each with its own characteristics and requirements. To systematically address these diverse needs, we have developed a formalism that categorizes MIV tasks based on their sources, templates, and targets. This formalism not only provides a common language for discussing MIV tasks but also helps in identifying patterns and best practices across different simulation domains.

In this formalism, we define \MIV as the act of synthesizing data from one or more different \textit{Sources} to dynamically generate a verification \textit{Template}, which defines the content pattern required to pass verification. This verification Template is in turn applied to an input file (the \textit{Target}) to perform the actual verification, returning a correct outcome if a match is achieved, and an error if not. Now the \MIV task can be performed in different ways, and we provide a simple formalism in Figure~\ref{fig:formalism} to help understand the different patterns that can be created.

Here, each pattern is described with a dot-delimited code, consisting of three components: the Source (or sources) using an Arabic numeral symbol, the Template Type using a capitalized letter symbol and the Target using a Roman numeral symbol. We provide two example pattern definitions in Figure~\ref{fig:formalism}. For instance, a \texttt{MIV 1.A.i} pattern could be a check that all locations in a geographical location file have a population of at least 0, while a pattern of type \texttt{MIV 4.C.ii} might (i) check whether the simulation is configured to explicitly model flooding events and then (ii) check whether locations in that same geographic file have, for example, an altitude and water holding capacity value defined if this is the case. 

Sources that may be used to generate the template may be content from the target file itself (1, as in our \texttt{MIV 1.A.i} example), from other input files (2), external reference information such as a lookup table or calendar (3) or simulation configuration files (4, as in our \texttt{MIV 4.C.ii} example). It can be possible that a \MIV pattern draws from multiple sources, such as the target file (1) and simulation configuration files (4). In this case the Arabic numerals can be appended in numerical order, giving the value ``14'' for the first component in this case. \MIV can be of different types, because they can be applied in different ways. These types include specifications that are statically applied to check a file (A, as in our \texttt{MIV 1.A.i} example), specifications that may be modified depending on specific criteria (B), specifications that may or may not be applied depending on specific criteria (C, as in our \texttt{MIV 4.C.ii} example), or (BC) specifications that may be modified or not be applied depending on specific criteria. Normally, \MIV of type A tends to be done either using only the target file as source (1.A.*) or the simulation configuration (4.A.*).

Lastly, \MIV patterns may differ in what aspect of the input file they verify, i.e. what they target. They may target for instance an individual column in a tabular data file (i, as in our \texttt{MIV 1.A.i} example), multiple static columns in a tabular data file (ii, as in our \texttt{MIV 4.C.ii} example), a dynamic number or arrangement of columns in a tabular data file (iii). There are also \MIV patterns that target files as a whole, and may target non-tabular model input files (v and higher). This may be done specifically to verify the syntax of the input file (vi), the nesting structure (vii, particularly useful for YAML-based input files) and the adherence to a predefined schema (viii, useful for both XML and YAML files for instance).

Given the three components and their variations, we are therefore able to define a total of at least 72 \MIV patterns, and more if we include patterns that rely on multiple Source types. However, the range of \MIV patterns is not intended to be exhaustive, and there are valuable input verification checks that we chose to leave outside of this formalism to retain simplicity. Most of these verification checks are checks that operate on 0 or multiple files, such as verification checks that operate on network-fed input data, checks that verify the number of input files present or checks that verify the non-existence of redundant or possible disruptive input files.
\subsection{\MIV in the context of SEAVEA}
Our \MIV tool can be applied to any application that requires input files in one of the supported formats. However, the benefits of the tool are amplified in cases where applications, and indeed even input files, are shared between users.

The SEAVEA project (Software Environment for Actionable VVUQ-enabled Exascale Applications, https://www.seavea-project.org), has established tools where this is the case. The toolkit itself provides facilities for the verification, validation and uncertainty quantification of HPC applications, and is an extension of the VECMA toolkit ~\cite{vecmatk}. For instance, within FabSim3~\cite{fabsim3} there are established plugins that contain sample input files for a range of different application domains. There are plugins available for applications in e.g. migration, Covid-19, the climate, materials and fusion domains. Here, our tool allows users to verify the input files present in the shared repository, and improve the quality of the input configurations for all other users.

The SEAVEA toolkit, and in particular FabSim3, also provides facilities to simplify the use of the MIV tool. For instance, FabSim3 enables external tools to be used through simple one-liner bash commands, automatically locating the relevant configuration files for the user’s application using its internal database. In addition, invocations of the MIV tool can be directly integrated into existing FabSim3 commands. Through this integration, users can choose to apply input file verification automatically for their own daily simulation workflows. Although such automated MIV checking introduces a performance overhead of several seconds, it ensures that any input files that the user requires are verified without additional human effort.
\section{Exemplars}
\label{sec:exemplars}
This section demonstrates the capabilities of the \MIV toolchain by going through  common input verification scenarios. To showcase the general nature of our tool, we present three exemplars on: (i) conflict-driven migration, (ii) disaster evacuation and (iii) disease spread.

These exemplars were selected to illustrate a range of input verification challenges commonly encountered in simulation modeling. They progress from basic data type checks to more complex multi-file validations and domain-specific constraints. By presenting these diverse scenarios, we aim to demonstrate FabGuard's capability in handling various types of input data, file formats, and validation requirements.
By presenting real-world applications, we demonstrate how the tool integrates into existing simulation workflows. These exemplars serve not only as proof of concept but also as guidance for potential users, illustrating how FabGuard can be adapted to different domains and specific verification needs.

%We focus on Agent-Based Models (ABMs) due to their diverse applications, complex input requirements, and sensitivity to input errors. ABMs provide an excellent testbed for FabGuard's capabilities, involving multiple, interconnected input files that describe agent characteristics, environment properties, and simulation parameters. While we use ABMs, FabGuard is a generic solution applicable to various simulation types and input file formats.
%These exemplars bridge the gap between theoretical concepts and practical implementation, illustrating FabGuard's potential to improve input verification across the broader landscape of computational modeling and simulation.

We chose to focus on Agent-Based Models (ABMs) for our exemplars due to their diverse applications across scientific disciplines, complex input requirements, and sensitivity to input errors. ABMs typically involve multiple, interconnected input files describing agent characteristics, environment properties, and simulation parameters, providing an excellent testbed for FabGuard's capabilities. Moreover, ABMs are often developed by researchers from diverse backgrounds, aligning with FabGuard's goal of making input verification more accessible to domain experts. While we concentrate on ABMs, it's important to note that FabGuard is a generic solution applicable to a wide range of simulation types and input file formats. The principles and techniques demonstrated in these exemplars can be readily adapted to other simulation paradigms, showcasing FabGuard's flexibility and potential to improve input verification across the broader landscape of computational modeling and simulation.

\subsection{Exemplar 1: Conflict-driven migration modelling with Flee}
As already mentioned in \S~\ref{sec:overview}, Flee~\cite{suleimenova2017} is a simulation tool designed for modeling displacement and migration patterns. It enables researchers to create simulations based on conflict and disaster scenarios, helping to predict how populations move in response to various crises. 
%Flee is in use across different institutions today to assist in migration modelling and it has been used in major research initiatives such as the EU-funded HiDALGO~\footnote{https://hidalgo-project.eu} and ITFLOWS~\footnote{https://itflows.eu} projects.

Flee models agents that move across a location graph: here, the location graph is defined using two input CSV files (locations.csv and routes.csv). Errors in the location graph input files not only lead to inaccuracies in the simulation, but can also lead to agents getting stuck in certain locations or to Flee to crash altogether. Another important input file for Flee version 3~\cite{ghorbani2024} is simsetting.yml, which is used to configure the set of assumptions used in the simulation. Lastly, there are a range of CSV files that define attributes for the spawned agents, as well as for specific locations and routes.

\begin{tcolorbox}[title=\MIV 1.A.i: Domain-specific constraints on a single column]
\begin{itemize}
\item \texttt{population} $>=$ 0
\item \texttt{location\_type} $\in$ [\texttt{"conflict\_zone", "town", "camp"]}
\end{itemize}
\end{tcolorbox}

The code snippet in Listing~\ref{lst:single} defines a schema for a \texttt{pandas DataFrame} using the \texttt{pandera} class \texttt{DataFrameModel}. It specifies that the \texttt{DataFrame} should have a \texttt{"population"} column with floating-point numbers greater than 0, which can also be null, and a \texttt{"location\_type"} column with string values that must be one of \texttt{"conflict\_zone," "town,"} or \texttt{"camp."} The \texttt{Check} function is used to enforce these constraints, with \texttt{Check.greater\_than(0)} ensuring the \texttt{"population"} values are positive and \texttt{Check.isin(["conflict\_zone," "town," "camp"])} ensuring the \texttt{"location\_type"} values are within the specified set. This schema validates the DataFrame's structure and data integrity by checking that the columns match the defined types and conditions.

\begin{lstlisting}[language=Python,caption = Single-column constraints, label={lst:single}]
class LocationsScheme(pa.DataFrameModel):
    location_type: Series[pa.String] = pa.Field(
        isin = ["conflict_zone", "town","camp"])
    population: Series[float] = pa.Field(ge=0,nullable=True,coerce=True)
\end{lstlisting}

We can refine the schema further as to accommodate domain-specific contraints that span multiple columns. 
\begin{tcolorbox}[title=\MIV 1.B.ii Multi-column constraint]
Locations that are conflict zones require a population value strictly higher than 0 (one needs persons to create conflict-driven displacement):
\end{tcolorbox}

The provided code snippet in Listing~\ref{lst:multicol} defines a custom validation function for a pandas DataFrame using the \texttt{pandera} library. The \texttt{@pa.dataframe\_check()} annotation designates the function \texttt{population\_gt\_0} as a custom DataFrame validation check. This function ensures that rows with \texttt{"location\_type"} equal to \texttt{"conflict\_zone"} do not have a \texttt{"population"} value less than or equal to 0. It creates a boolean mask to identify these invalid rows and raises a \texttt{ValueError} with the indices of any invalid rows found. The function then returns a boolean Series indicating which rows are valid. By using the \texttt{@pa.dataframe\_check()} annotation, this custom check is integrated into a \texttt{pandera} schema, allowing it to be used in the validation process to enforce specific data constraints.

\begin{lstlisting}[language=Python, caption = Multi-column constraints, label=lst:multicol]
@pa.dataframe_check()
def population_gt_0(cls, df: pd.DataFrame) -> Series[bool]:
    # Define conditions based on 'location_type' and 'population' columns
    mask = ((df["location_type"] == "conflict_zone") & (df["population"] <= 0))

    # Filter the DataFrame to keep only valid rows
    if mask.any():  # Check if any rows meet the condition
        # Print the rowa that do not meet the condition
        raise ValueError(df.index[mask])
    return ~mask
\end{lstlisting}

\begin{comment}
\subsubsection{Multi-column constraint scenario: Flee locations} 
As an extension to the example above, our MIV tool is also able to express constraints that have dependencies between columns. For instance, we could enforce that locations that are conflict zones require a population value higher than 0 (one needs persons to create conflict-driven displacement. This would be represented as follows:

%\lstinline|if location_type == "conflict\_zone" then population > 0"|

\begin{lstlisting}[language=Python]
schema = pa.DataFrameSchema(
{
"population": pa.Column(float, [
    pa.Check( lambda g: g["conflict_zone"] > 0, groupby=["location_type"])]
})
\end{lstlisting}
\end{comment}

\begin{tcolorbox}[title=\MIV 2.A.ii Constraints spanning multiple files]
Within Flee, the countries featured in the model are located in locations.csv, but any border closures are defined in closures.csv. We must ensure closures link to the correct countries (and for instance do not have typos in the country names).
\end{tcolorbox}

We can apply the same ideas as above: create a boolean mask that identified the invalid rows and raise an errors if such entries are found. One caveat in comparison to the previous example is that we need to load the locations.csv file. The final constrints is implemented in Listing~\ref{lst:multi}. 

\begin{lstlisting}[language=Python,caption = Constraints across files, label=lst:multi]
@pa.dataframe_check()
def closure_type_country(cls, df: pd.DataFrame) -> Series[bool]:
    dfl = # Load the content of the "locations" file
    # Get a list of countries from the "locations" file
    loc_countries = dfl["country"].tolist()

    # Define a mask to check if the conditions are met
    mask = ((df["closure_type"] == "country")
            & (~df["name1"].isin(loc_countries)
            & (~df["name2"].isin(loc_countries))))

    # ... raise an errors or return the valid entries
\end{lstlisting}

\begin{comment}
\subsubsection{Constraints spanning multiple files: verified country closures}
Within Flee, the countries featured in the model are located in locations.csv, but any border closures are defined in closures.csv. To ensure that these border closures indeed link to the correct countries (and for instance do not have typos in the country names), we can introduce the following verification check:

\begin{lstlisting}[language=Python]
ADD EXAMPLE
if closure_type=="country" then closure.name1 in location.country
\end{lstlisting}
\end{comment}

\subsection{Exemplar 2: Disaster-driven evacuation modelling with DFlee}
Dflee~\cite{jahani2023} is a variation of Flee which is configured to model disaster-driven population displacement. The simulation tool currently is used for flood-driven migration, but extensions to capture other events (such as storms) are in progress.

Like Flee, DFlee relies on a location graph, but depending on the context the location and route attributed may be radically different. Errors in these input files may result in problems similar to Flee, or in a complete lack of spawned agents in the simulations. DFlee also relies on a simsetting.yml, and a number of fields in there need to be defined correctly for the DFlee to be triggered, while other values need to be lined up in a consistent manner to allow DFlee to work in a manner that matches basic logic (e.g. that people are more likely to flee from flooded areas than unflooded ones). When used for flooding, DFlee also requires a flood\_level.csv file, which contains the progression of the flooding at each location during the simulation period. Errors within this file may cause flooding to occur at the wrong times, in the wrong places, or with the wrong intensities.

\begin{tcolorbox}[title=\MIV 3.A.i Custom-function columns constraints]
Validating that a day column has valid rows for all days in a month
\end{tcolorbox}
The code in Listing~\ref{lst:stepwise} defines a custom check function \texttt{check\_day\_increment} using the \texttt{pandera} library, annotated with \texttt{@pa.check("Day")} to specify that it applies to the "Day" column of a DataFrame. The function validates that the values in the "Day" column are incremental integers within a specified range. It sets a minimum value of 0, a maximum value determined by reading the configuration file, and a step increment of 1. The function returns a boolean Series indicating whether each value in the "Day" column meets these conditions: being an increment of 1 from the minimum value, and lying within the inclusive range from the minimum to the maximum value. This ensures that the "Day" column contains valid, sequential day values.

\begin{lstlisting}[language=Python, caption = Stepside checks, label=lst:stepwise]
@pa.check("Day")
def check_day_increment(cls, series: Series[int]) -> Series[bool]:
    min_value = 0  # define your min value
    max_value = get_sim_period_len()  # define your max value
    step = 1  # define the step increment

    # Check if each value is an increment of `step` within the range [min_value, max_value]
    return ((series - min_value) % step == 0) & (series >= min_value) & (series <= max_value)
\end{lstlisting}

\begin{tcolorbox}[title=\MIV 4.C.iii Dynamic columns constraints]
When used for flooding, DFlee also requires a flood level.csv file, which contains the progression of the flooding at each location during the simulation period. Errors
within this file may cause flooding to occur at the wrong times, in the wrong places, or with the wrong intensities
\end{tcolorbox}

%In DFlee, all flood zone columns (in flood level.csv) should be within the max flood level defined in the sym- settings file. Similarly, the number of rows depend on a config parameter

Listing~\ref{lst:dynamic} demonstrate  another pattern which allows for dynamic schema validation where the same constraints should be applied to a varied number of columns. In the schema defined below, the number of columns in the flood level.csv is unknown, but all columns except the first specify the same type of information -- the intensity of the flood for each day for different flooz zones. where the rows are the days, and the columns are the flood zones. To realise these constraints, we have defines a class method \texttt{with\_dynamic\_columns} within a \texttt{FloodLevelScheme} class that dynamically creates schema constraints for a pandas DataFrame. The method reads configuration values to set maximum permissible values for the "Day" and other flood levels columns. It generates fields with these constraints, and specifying value ranges for all columns. These constraints are added to a dictionary and used to create a new class, \texttt{ExtendedFloodLevelScheme}, which inherits from \texttt{FloodLevelScheme} and includes the dynamically generated attributes. 

\newpage

\begin{lstlisting}[language=Python, caption = Schema with dynamic columns,label=lst:dynamic]
class FloodLevelScheme 
    @classmethod
    def with_dynamic_columns(cls, df: pd.DataFrame):
        flood_zone_max_value = #read from config
        #Create contraints for the day column: value range of 0 to day_max_value
        
        # Add constraints for all but the first columns
        for column in df.columns[1:]:
            ...
            all_other_fields = pa.Field(...
                in_range={"min_value": 0, "max_value": flood_zone_max_value})
            dynamic_attrs[column] = all_other_fields
    
        # Create a new dynamic class with columns as defined in dynamic_attrs
        return type('ExtendedFloodLevelScheme', (FloodLevelScheme,), dynamic_attrs)
\end{lstlisting}

\subsection{Exemplar 3: Disease spread modelling with FACS}
FACS (Flu And Coronavirus Simulator)~\cite{mahmood2022} is a computational modeling tool designed to simulate the spread of influenza and coronaviruses such as COVID-19 in various populations and settings. It allows users to explore the impact of different public health interventions, such as social distancing, vaccination, and lockdown measures, on the spread of these infectious diseases. 

%\begin{tcolorbox}[mybox, title= MIV 3.A.i Add example on calendar verifications for measures.yml]
%TODO: add description, code etc.
%\end{tcolorbox}

To configure individual simulations, FACS relies in a wide range of input files. These include input files to provide geographical information (buildings.csv), demographic information (age-distr.csv and needs.csv), disease information (e.g. disease\_covid19.yml and mutations.yml) as well as information on interventions (measures.yml) and vaccination types and strategies (vaccinations.yml). Users commonly edit the measures.yml file to assess the efficacy of new intervention scenarios, and this file is relatively complex in terms of structure. Erroneous entries in measures.yml can have wide-ranging results. For instance, interventions may not trigger at all or they may trigger with the wrong intensity.

\begin{tcolorbox}[title= \MIV 3.A.viii Schema-based summation check]
%\textbf{Scenario 4: The sum of all but the first columns should be 1} Many data files contain percentage information. The first challenge when you work with general constraints that can be applied to files with different dimensions is that the columns can be added dynamically.
All demographic files (e.g. demographic\_age, demographic\_gender, etc) for FACS and DFlee contains columns which lists representative fractions of the population. Respectively, the sum of all entries in these columns should add up to 1 (the number required could be modified for different use cases). 
%that can express the constraint across all demographic type files. (demographic\_age, demograohic\_gender, etc) 
\end{tcolorbox}

Listing~\ref{lst:multiple} implements a \texttt{DemographicScheme} class, which inherits from \texttt{pa.DataFrameModel} in the \texttt{pandera} library, includes a custom validation method \texttt{all\_but\_first\_column\_sum\_is\_1} marked with the \texttt{@pa.dataframe\_check} decorator. This method ensures that the sum of the values in all columns, except the first one, equals 1. It iterates through each column (excluding the first), calculates the sum of its values, and checks if it equals 1. If any column's sum is not equal to 1, it appends the column name and its sum to an \texttt{errors} list. If there are errors, the function would report them; otherwise, it returns \texttt{True}, indicating the DataFrame meets the validation criteria. 

\begin{lstlisting}[language=Python, caption= Schema across Multiple files, label={lst:multiple}]
class DemographicScheme(pa.DataFrameModel):
    # name: Series[pa.String] = pa.Field(nullable=False, alias='#"name"')

    @pa.dataframe_check
    def all_but_first_column_sum_is_1(cls, df: DataFrame) -> bool:
        # Iterate over the names of all columns except the first one
        errors = []
        for column_name in df.columns[1:]:
            column_sum = df[column_name].sum()
            if column_sum != 1:
                errors.append(f"{column_name},{column_sum}")
        if len(errors) > 0:
           # Report the errors 
        return True
\end{lstlisting}

\begin{tcolorbox}[title=\MIV 1.A.vii Nested entries yaml validation]
In addition to having the correct types, yaml entries should be  correctly indented as to preserve the intended meaning. For example, the partial\_closure section in the measures.yml allows nested entries, such as for shopping centers, hospitals, etc., enabling detailed specifications for various facilities. 
\end{tcolorbox}

A key insight in our FACS verification journey was that the majority of the FACS yaml verification requirements could be met through off-the-shelf schema validation. Capitalizing on YAML's compatibility as a superset of JSON, we utilized a well-known Python library designed for JSON schema validation. This schema not only specifies the types for each data entry but also outlines the structure, including the hierarchy of entries and the allowance for nested entries. 

%Writing the schema itself is the most challenging aspect of this process. However, our preliminary tests have shown that LLMs are highly effective at generating schemas, having successfully produced initial schemas for all of FACS' input files.

An excerpt from the jsonschema for the measures.yaml file is given below:  

\begin{lstlisting}[language=JSON, caption= Json Schema, label={lst:jsonschema}]

 "partial_closure": {
    "type": "object",
    "properties": {
     "leisure": {"type": "number", "minimum": 0, "maximum": 1},
      "school": {"type": "number", "minimum": 0, "maximum": 1},
      "shopping": {"type": "number", "minimum": 0, "maximum": 1},
      "example": {"type": "number", "minimum": 0, "maximum": 1}
    },
    "additionalProperties": false ...
\end{lstlisting}

This JSON schema implements the requirements for correctly indented YAML entries with nested structures in the $"partial\_closure"$ section. It defines $"partial\_closure"$ as an object with specific properties (e.g., "leisure", "school") as numbers between 0 and 1. With 
"additionalProperties" set to false, it strictly limits entries to these predefined types. This ensures a YAML structure where $"partial\_closure"$ is the main section, with only the specified facility types indented beneath it, directly translating the schema's hierarchy into proper YAML indentation and preserving the intended nested relationship.

Through these exemplars, we demonstrate how FabGuard can handle a variety of input verification scenarios, from simple data type checks to complex multi-file validations and domain-specific constraints. This range of examples illustrates the tool's potential to enhance the reliability and reproducibility of simulations across different scientific domains.

\section{LLMs for Constraints Inference and Generation}
\label{sec:llm}

The adoption of Model Input Verification (MIV) practices faces challenges due to the complexity of setting up verification frameworks and the need for domain-specific knowledge. To address these usability concerns and lower the barrier to entry for MIV, we explored the potential of Large Language Models (LLMs) in automating parts of the MIV process. LLMs, with their ability to understand and generate human-like text, offer a promising approach to inferring constraints from existing data and generating new constraints based on natural language descriptions. This section investigates two key research questions:

\begin{enumerate}
    \item RQ1: Can LLMs be used for constraints inference?
    \item RQ2: Can LLMs be used for constraints generation?
\end{enumerate}

By leveraging LLMs, we aim to make MIV more accessible to simulation practitioners who may not have extensive programming backgrounds or in-depth knowledge of data validation techniques.

\subsection{RQ1: Constraints Inference}

To address RQ1, we conducted an experiment using To address RQ1, we conducted an experiment using Claude 3.5 Sonnet \footnote{claude.ai}, a language model developed by Anthropic \footnote{https://www.anthropic.com/} and released in 2024. Claude's ability to understand and generate code makes it suitable for our constraint inference experiment. We provided Claude with input files for the Flee simulation, along with explanations of the simulations and instructions on using Pandera for validation.

\textbf{Methodology:} Our approach involved several key steps. First, we supplied Claude with the contents of key input files, including locations.csv, routes.csv, and closures.csv for Flee.
%, and additional files like flood\_level.csv for DFlee. 
We then provided detailed explanations of the simulation, including the purpose of each input file. 
%and the relationships between them. 
We introduced Claude to Pandera, explaining its use for DataFrame validation and providing examples of how to create schemas and custom checks. Finally, we asked Claude to infer and generate Pandera schemas and checks based on the provided information.

\textbf{Findings:} Table~\ref{tab:constraint-comparison} presents a comparison of key constraints inferred by Claude against our manual tests. We categorized the constraints into four types: simple single-column, refined single-column, multi-column, and multi-file. Simple single-column constraints, which only specify column data types, are omitted from the table. Flee contained 12 such constraints across its three input files. Claude precisely inferred 10 of these and enhanced two date-related single-column constraints (represented as integers) by adding a "greater than 0" restriction. Refined single-column constraints involve validations beyond simple data types, such as ranges or set memberships. Multi-column and multi-file constraints involve relationships between multiple columns or files, respectively. %This categorization allows for a more nuanced analysis of Claude's inference capabilities across different levels of constraint complexity.

Our experiment revealed that Claude was capable of inferring a wide range of constraints, including some that were not present in our manual tests. In the analysis of Flee's constraints, 22 out of 23 constraints were correctly inferred, with no wrong inferences. Specifically, 17 constraints were precisely inferred, while one constraint was not inferred at all. Two constraints were corrected from their initial incorrect state, namely the longitude range and route checking. Another two constraints were improved and made stronger than initially proposed. Lastly, one constraint was inferred but was weaker than the actual constraint. This analysis suggests that the inference process aligns closely with the constraints generated by an expert (the second author) working on the tool, though some adjustments were needed to fully capture all aspects of the constraints. 

%For DFlee, 6 out of 7 constraints were correctly inferred. 

\begin{table}[htbp]
\caption{Comparison of Constraints in Manual Tests vs. LLM-Inferred Tests}
\label{tab:constraint-comparison}
%\resizebox{\textwidth}{!}{%
\footnotesize
\renewcommand{\arraystretch}{1.2}  % Increase row height
\begin{tabular}{|c|p{0.28\textwidth}|p{0.28\textwidth}|p{0.18\textwidth}|}
\hline
\textbf{Category} & \textbf{Manual Test} & \textbf{LLM-Inferred Test} & \textbf{Status} \\
\hline
\multirow{5}{*}{\rotatebox[origin=c]{90}{\centering\textbf{Single-column \hspace{7mm}}}} 
 & Coordinates within [-180, 180] & Latitude [-90, 90], Longitude [-180, 180] & Improved(Corrected) \\
\cline{2-4}
 & Location type in ["conflict\_zone", ..., "marker", "idpcamp"] & Location type in ["conflict\_zone", "town", ...] & Partial (missing "marker" and "idpcamp") \\
\cline{2-4}
 & Route distance $>$ 0 & Route distance $\geq$ 0 & Improved(Corrected) \\
\cline{2-4}
 & Forced redirection in [0, 1, 2] & Forced redirection in [0, 1, 2] & Exact match \\
\cline{2-4}
 & Closure type in ["location", "country", "links", "camp", "idpcamp"] & Closure type in ["country", "camp"] & Partial (missing "location", "links", "idpcamp") \\
\hline
\multirow{4}{*}{\rotatebox[origin=c]{90}{\textbf{Multi-column} \hspace{7mm}}} 
 & Population $>$ 0 for camp, town, conflict; = 0 for markers; $\geq$ 0 for forwarding hub & Population $\geq$ 0 for all location types & Requires adjustment (less specific) \\
\cline{2-4}
 & Conflict zones must have a conflict date & Conflict zones must have a conflict date & Exact match \\
\cline{2-4}
 & First country in country column applies to all conflict zones & -- & Not inferred \\
\cline{2-4}
 & Location names must be unique & Location names must be unique and non-null & Match (Enhanced) \\
\hline
\multirow{2}{*}{\rotatebox[origin=c]{90}{\textbf{Multi-file} \hspace{1mm}}} 
 & Closure countries (name1, name2) must be valid countries from locations file & Implemented cross-file check for valid countries in closures & Exact Match \\
\cline{2-4}
 & Location names must exist in routes file (as name1 or name2) & Suggested cross-file check for location names in routes & Exact Match \\
\hline
\end{tabular}
%}
\end{table}

\begin{comment}
Claude successfully corrected some constraints, it generated a check for the geographical coordinates in the Flee locations file:

\begin{lstlisting}[language=Python]
latitude: Series[Float] = pa.Field(ge=-90, le=90)
longitude: Series[Float] = pa.Field(ge=-180, le=180)
\end{lstlisting}

This constraint ensures that the latitude and longitude values are within valid ranges, which is important for the correct geographical representation in the simulation. In our initial tests, both latitude and longitude were wrongly set between -180 and 180. 
\end{comment}

Claude generated several constraints absent from manual tests. For routes.csv, it introduced checks for distinct route endpoints, unique location names, and prevention of duplicate routes. In closure.csv, it validated that end dates should be after the start dates and that the non-null value of a column (name2) depends on another column (closure type). Claude also developed two multi-file constraints: ensuring camp closures reference valid camps from the locations file, and identifying isolated locations. The latter was implemented as:

\begin{lstlisting}[language=Python]
# Constraint: Check for isolated locations (not connected by any route)
connected_locations = set(routes_df['name1']) | set(routes_df['name2'])
isolated_locations = set(locations_df['name']) - connected_locations
if isolated_locations:
    print(f"Warning: The following locations are isolated (not connected by any route): {isolated_locations}")
\end{lstlisting}

This check can reveal potential data errors or geographical inconsistencies in the simulation. These AI-generated constraints demonstrate Claude's ability to infer validation rules addressing data integrity, consistency, and cross-file relationships in the Flee system, potentially identifying errors overlooked in manual testing.

%This check ensures that the probabilities in demographic files sum to 1, which is essential for maintaining the integrity of the population distribution in the simulation.

%However, there were also cases where Claude's inferences differed from our manual tests. For instance, in the DFlee flood level schema, Claude assumed normalized flood levels between 0 and 1, while our actual implementation allows for a configurable maximum flood level. This difference highlights the importance of providing specific configuration details to the LLM for more accurate constraint inference.

\textbf{Implications:} %These findings have several implications for the use of LLMs in MIV. 
LLMs can effectively infer a wide range of constraints, potentially accelerating the initial stages of MIV development. They can complement manual tests by identifying additional checks that human developers might overlook. However, the accuracy of LLM-inferred constraints can be improved by providing more detailed configuration information, and data.

\subsection{RQ2: Constraints Generation}

For RQ2, we explored Claude's ability to generate specific constraints when provided with clear descriptions of the constraint.

\textbf{Methodology:} Our approach involved providing Claude with detailed descriptions of constraints, using the same format as in the Exemplars section of this paper. We then asked Claude to implement these constraints using Pandera, specifying that the implementation should include necessary imports and class structures. Finally, we manually reviewed the generated code to assess its correctness and completeness in implementing the described constraints.

\textbf{Findings:} Claude demonstrated a high degree of accuracy in generating constraints based on descriptions. Out of 13 constraint descriptions provided, Claude successfully generated 11 correct implementations. Of the remaining two, both required minor adjustments. Table \ref{tab:constraint-generation} presents examples of constraint descriptions and Claude's implementations. In both cases, Claude accurately translated the constraint descriptions into functional Pandera checks. The generated code not only implements the logical constraints but also follows Pandera's syntax and best practices. However, we observed that for more complex constraints, especially those involving configuration-dependent values or specific simulation logic, Claude's implementations required minor adjustments. For example, in constraints involving maximum flood levels in DFlee, Claude initially used hard-coded values, which we needed to replace with configuration-dependent variables. 

\begin{table}[htbp]
\caption{Examples of Constraint Descriptions and Generated Implementations}
\label{tab:constraint-generation}
\begin{tabular}{|p{0.45\textwidth}|p{0.45\textwidth}|}
\hline
\textbf{Constraint Description} & \textbf{Generated Implementation} \\
\hline
"Route distances must be positive numbers" & 
\begin{lstlisting}[language=Python]
distance: Series[Float] = pa.Field(gt=0)
\end{lstlisting} \\
\hline
"The sum of all entries in demographic probability columns should add up to 1" & 
\begin{lstlisting}[language=Python]
@pa.dataframe_check
def probabilities_sum_to_one(cls, df: pd.DataFrame) -> bool:
    prob_columns = [col for col in df.columns if col != 'category']
    return all(df[prob_columns].sum(axis=1).between(0.99, 1.01))
\end{lstlisting} \\
\hline
\end{tabular}
\end{table}

%\textbf{TODO: Give example from Flee and show the one requireing adjustments}

\textbf{Implications:} %The findings from RQ2 suggest several implications. 
LLMs can significantly speed up the initial implementation of MIV constraints, particularly for common validation patterns and clearly described requirements. The accuracy of constraint generation for well-defined rules suggests that LLMs could be valuable tools in translating natural language specifications into code.%, potentially bridging the gap between domain experts and developers. 
While LLMs excel at generating syntactically correct and logically sound constraints, there's still a need for domain expert review, especially for constraints involving simulation-specific or configuration-dependent logic. %The ability of LLMs to generate constraints based on descriptions opens up possibilities for more interactive and iterative MIV development processes, where domain experts can refine constraints through natural language interactions with the LLM.

\subsection{On the potential use of LLMs in MIV}
Our experiments with Claude on the Flee case study demonstrate that LLMs have significant potential in both inferring and generating constraints for Model Input Verification. They excel at identifying a wide range of constraints and can accurately translate natural language descriptions into functional code.

%This capability is particularly important in the context of simulation modelling, which is often developed in environments quite different from traditional software engineering. Typically, these simulations are written by domain experts - scientists, researchers, and analysts - who, while highly skilled in their fields, may not have extensive programming backgrounds. As a result, complex coding patterns, unit testing, and other software engineering best practices that are commonplace in the development community are not as widely adopted or implemented in the simulation contexts.
This capability is especially important in simulation modeling, often developed by domain experts who may lack extensive programming backgrounds. 
%The gap between domain expertise and software engineering practices has long been a challenge in ensuring the reliability and verifiability of scientific simulations. 
Our preliminary analysis shows that LLMs, when provided with the right setup - including appropriate structure, classes, and examples - can bridge the gap between domain expertise and software engineering practices, at least in the context of input verification. They make the process of writing constraints more accessible and bring the power of formal specification to domain experts who may not have deep programming knowledge. 

While LLMs show promise in MIV, their use presents challenges. Our experience revealed a shift from quick constraint generation to time-consuming validation, emphasizing the need for human expertise.
Generating constraints with LLMs was quick, taking less than an hour, but validating their accuracy required a several hours of work. LLM-generated constraints, though technically correct, often proved overly conservative, missing potential valid types not present in sample data. This highlights the importance of comprehensive datasets and domain expert involvement when using LLMs for constraint generation. While LLMs can accelerate initial constraint generation, they complement rather than replace human expertise in the MIV process.

%The experiment also highlight areas for improvement, such as the need for more detailed configuration information to avoid assumptions about data ranges. The results suggest that LLMs could be valuable tools in MIV workflows, potentially accelerating development and improving the comprehensiveness of input validation. Future work should focus on developing specialized prompts or fine-tuned models to enhance the accuracy of constraint inference and generation, particularly for domain-specific and configuration-dependent constraints.
\section{Evaluation}
\label{sec:evaluation}

\begin{figure}[t!]
\centering
\begin{subfigure}[b]{0.48\textwidth}
    \includegraphics[width=\textwidth]{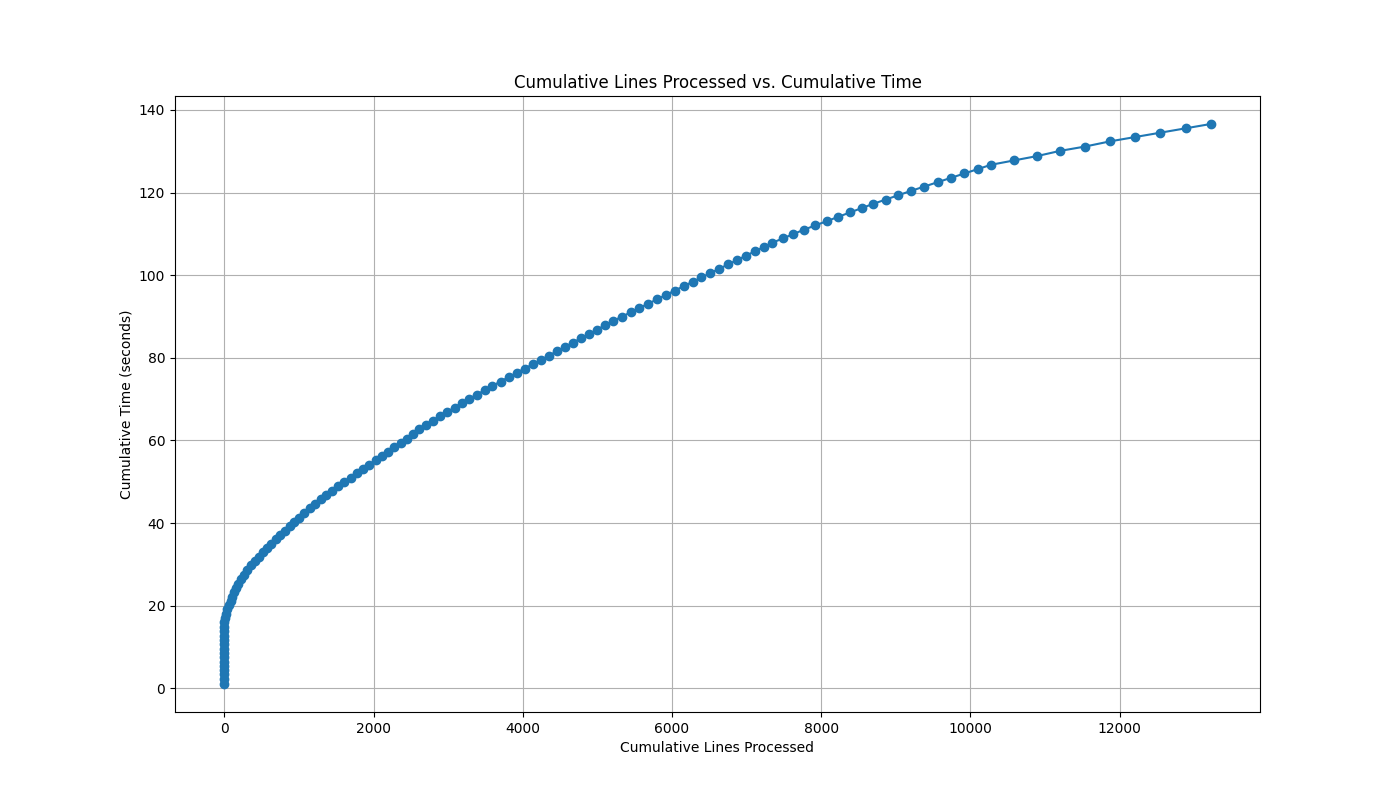}
    \caption{Cumulative lines vs. cumulative time}
    \label{fig:cumulative}
\end{subfigure}
\hfill
\begin{subfigure}[b]{0.48\textwidth}
    \includegraphics[width=\textwidth]{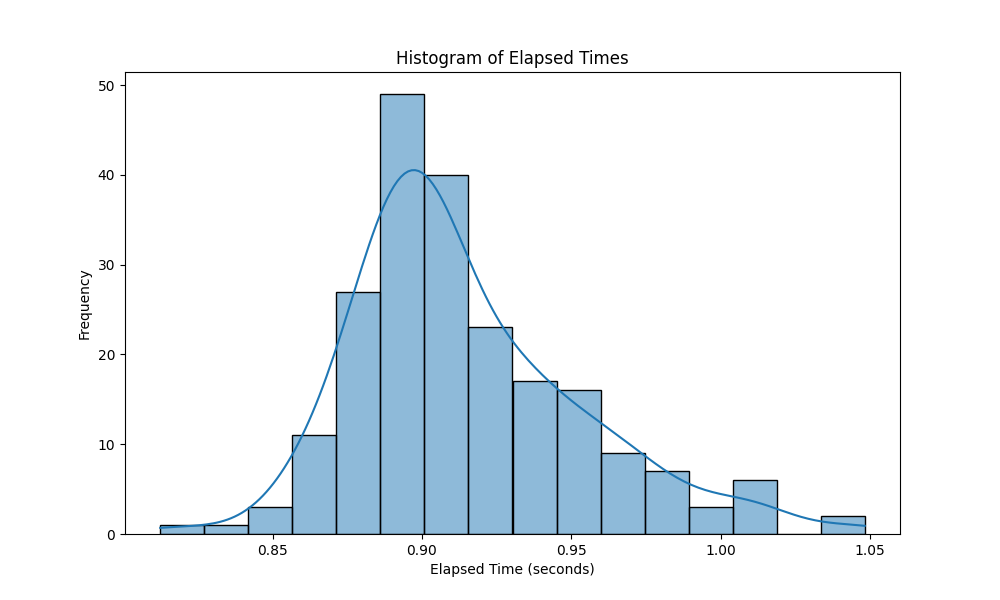}
    \caption{Elapsed time histogram}
    \label{fig:histogram}
\end{subfigure}

\begin{subfigure}[b]{0.48\textwidth}
    \includegraphics[width=\textwidth]{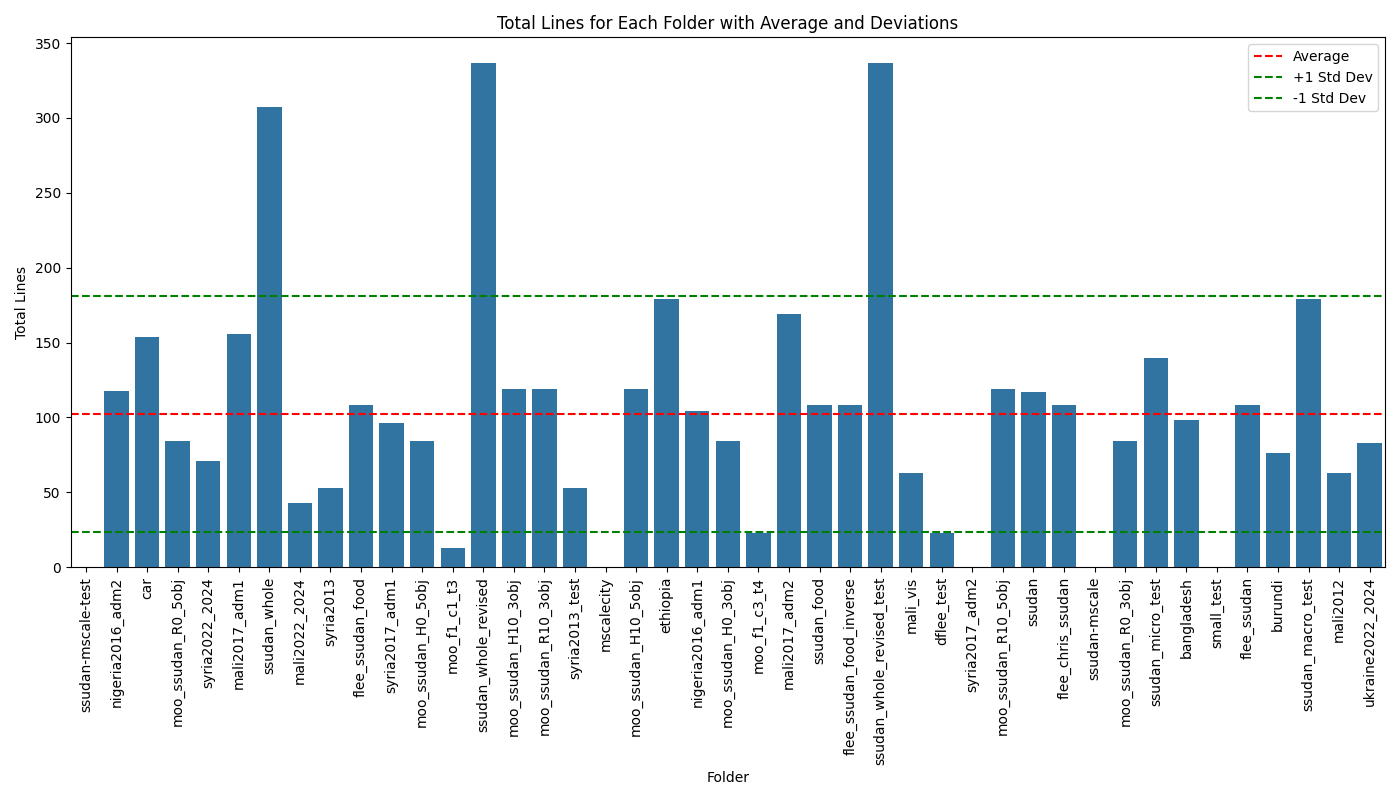}
    \caption{Total lines per conflict}
    \label{fig:total_lines}
\end{subfigure}
\hfill
\begin{subfigure}[b]{0.48\textwidth}
    \includegraphics[width=\textwidth]{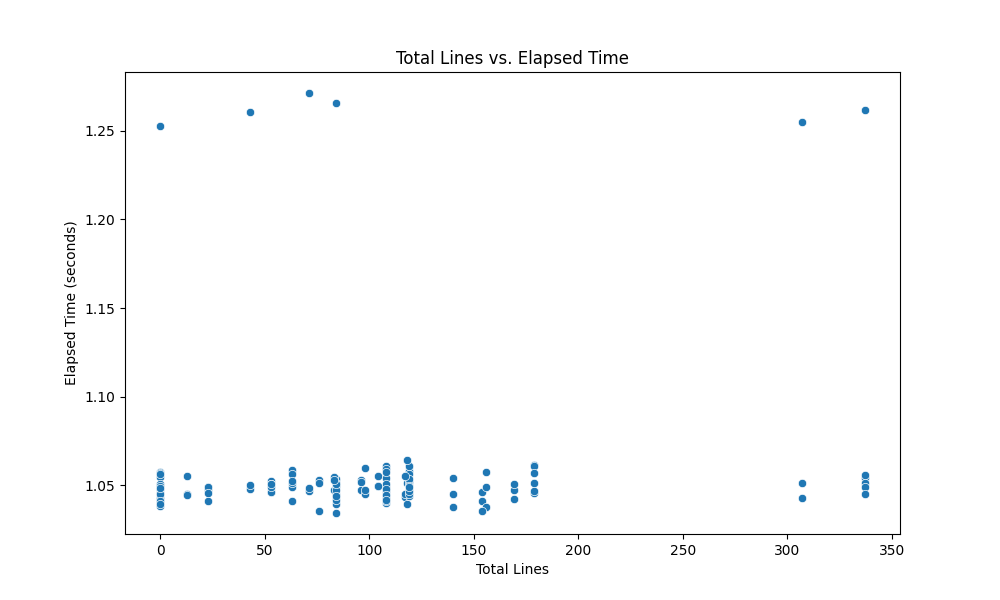}
    \caption{Total lines vs. elapsed time}
    \label{fig:scatter}
\end{subfigure}

\caption{Performance analysis of FabGuard on the Flee dataset}
\label{fig:flee_benchmarks}
\end{figure}
Our evaluation of FabGuard aims to demonstrate its scalability and applicability. We conducted two sets of tests: (1) Microbenchmarks with generated input files and tests; (2) a real-world simulation using the Flee system and custom test files. Section~\ref{sec:microbench} presents the microbenchmark results, while Section~\ref{lbl:usecase} shows the results on FLEE. These tests provide insights into FabGuard's performance across various scenarios, from controlled environments to practical applications.

\mypar{Setup.} Our evaluation was conducted on an Apple M2 Max with 12‑core CPU, 30‑core GPU and 16‑core Neural Engine, 64 GB of RAM, and 1TB of HDD running MacOS Ventura 13.5. We used Python 3.12.0. We repeat each benchmark for 5 warmup times and 30 execution times and report the average execution time.

\subsection{Use case: Flee}
\label{lbl:usecase}

We evaluate FabGuard's performance by running our test suite on the entire Flee conflicts dataset. The test suite consists of 4 Pandera files, each containing a specific number of constraints. This comprehensive evaluation covered 100 conflicts, encompassing a total of 12,000 input files. This  dataset allows us to assess FabGuard's efficiency and scalability across a wide range of real-world scenarios.

\begin{comment}
\begin{figure}[h]
\centering
\includegraphics[width=0.48\textwidth]{figures/fleebenchmarks/cumulative_lines_vs_cumulative_time.png}
\includegraphics[width=0.48\textwidth]{figures/fleebenchmarks/elapsed_time_histogram.png}
\includegraphics[width=0.48\textwidth]{figures/fleebenchmarks/total_lines_barchart.png}
\includegraphics[width=0.48\textwidth]{figures/fleebenchmarks/total_lines_vs_elapsed_time.png}
\caption{Figure}
\end{figure}
\end{comment}

The results of our evaluation are summarized in Figures 1-4, each highlighting different aspects of FabGuard's performance. Upon analyzing these results, several key insights emerge which we have summarised below.

\mypar{Scalability.} FabGuard demonstrates good overall scalability, processing approximately 12,000 lines in about 140 seconds (Figure~\ref{fig:cumulative}). 

%Overall, the smooth progression of the curve implies consistent performance across different input sizes, with the tool maintaining reasonable efficiency even as the codebase expands significantly.
%However, the non-linear curve suggests that processing efficiency slightly decreases as the number of lines increases, possibly due to increasing complexity or memory management overhead.

\mypar{Consistency.} The majority of files are processed within a narrow time range of 0.85 to 1.05 seconds, with a peak around 0.90 seconds (Figure~\ref{fig:histogram}). This consistency across different file sizes indicates a reliable performance baseline for FabGuard.

\mypar{Processing Time vs. File Size.} Interestingly, there isn't a strong linear relationship between file size and processing time for most files (Figure~\ref{fig:scatter}). 
This suggests that FabGuard has a relatively constant overhead for each file, with the actual content verification time being comparatively small. Moreoer, the Flee dataset exhibits significant variability in file sizes across different folders (Figure~\ref{fig:histogram}). Most folders contain files with fewer than 200 lines, but some exceed 300 lines. Despite this variability, FabGuard maintains relatively consistent processing times. A few files with longer processing times (1.25-1.27 seconds) create a slight right skew in the distribution (Figure~\ref{fig:histogram}), suggesting factors beyond line count can affect processing time.
\mypar{Efficiency.} Based on the overall processing of 12,000 lines in 140 seconds, FabGuard achieves an average processing rate of approximately 85.71 lines per second. This rate demonstrates FabGuard's efficiency in handling large datasets. FabGuard ability to process a large number of files quickly makes it suitable for real-world applications where rapid input verification and makes it a viable part of a CI/CD pipeline. %However, the slight decrease in efficiency with larger files and the presence of outliers suggest potential areas for future optimization, particularly in handling more complex or larger input files.

\subsection{Microbenchmarks}
\label{sec:microbench}

We designed a series of microbenchmarks aimed at stress-testing our approach under various conditions. These microbenchmarks evaluated FabGuard's behavior across four key dimensions: data types (1-10), number of columns (10-100) and rows (100-1000) per file, and total number of files processed (1-100). The experiments utilized randomly generated tests and files to simulate scenarios FabGuard might encounter in real-world applications. Results revealed consistent performance across these input dimensions, with no significant bottlenecks or scalability issues. While slight fluctuations in execution time were observed with changes in data complexity and file structure, these variations were minimal, typically within a range of 0.05 to 0.1 seconds (approximately 1-2\% of total execution time). The most notable finding was a linear correlation between the number of files processed and execution time, indicating predictable scaling for large-scale simulations.  %The absence of major performance bottlenecks across the tested parameters indicates that FabGuard can maintain consistent performance across a wide range of input scenarios. %Overall, FabGuard demonstrated robust performance and consistent memory usage across the wide range of test scenarios, suggesting its suitability for diverse simulation environments.

\section{Conclusion and Future Work}
\label{sec:conclusion}
In this paper, we have introduced Model Input Verification (MIV) as an essential step in enhancing the reliability and reproducibility of simulation models. Our primary contribution is a comprehensive methodology for MIV, implemented in the FabGuard toolset. This methodology adapts established data schema and validation tools to address the unique challenges of simulation input verification. We formalized MIV patterns, categorizing verification tasks based on their sources, template types, and targets. This formalism provides a structured approach to identifying and implementing input verification requirements across diverse simulation domains.

Our work goes beyond theoretical frameworks by demonstrating the practical application of these MIV patterns. We presented numerous examples across three domains: conflict-driven migration, disaster evacuation, and disease spread modeling. These case studies showcase how FabGuard can handle a variety of validation scenarios, from simple data type checks to complex multi-file validations and domain-specific constraints. Furthermore, we conducted the first study on using Large Language Models (LLMs) for constraint discovery and generation in the context of MIV. Our results show that LLMs can accurately infer existing constraints and even identify new, valid constraints, potentially lowering the barrier to entry for adopting robust MIV practices. This exploration of LLMs, combined with our identified requirements for MIV tools, establishes a foundational framework for the future development of model input verification systems. Our evaluation provided empirical evidence of MIV's feasibility for large-scale simulations, with FabGuard efficiently processing 12,000 input files in 140 seconds while maintaining consistent performance across varying file sizes and complexities.

These contributions establish a foundation for more robust and trustworthy simulation practices. We envision MIV becoming an integral part of the simulation modeling workflow, akin to unit testing in software development. Future research will focus on expanding FabGuard's capabilities to cover a broader range of simulation paradigms and input formats. We plan to conduct large-scale studies on the use of Large Language Models, for automated constraint discovery in complex, domain-specific relationships. This research will aim to further lower the barrier for MIV adoption and improve its effectiveness across diverse simulation domains. We will work on developing user-friendly interfaces to make MIV more accessible to non-technical users, bridging the gap between domain expertise and software engineering practices. We will further explore the integration of MIV with other stages of the simulation lifecycle, such as output validation and uncertainty quantification. This holistic approach could lead to a more robust framework for ensuring simulation reliability. Additionally, we will undertake case studies across diverse scientific domains to refine and validate MIV methodologies, providing empirical evidence of their effectiveness and generalizability.

This research contributes to establishing input verification as a fundamental component of the simulation modeling process, rather than an afterthought. By integrating MIV into standard modeling practices, we aim to enhance the reliability of simulations and, consequently, the quality of scientific discoveries based on these models. The broader adoption of systematic input verification techniques has the potential to improve the overall robustness and credibility of simulation-based research across various scientific disciplines.

%Future work will focus on exploring the integration of FabGuard with other stages of the simulation lifecycle, such as output validation and uncertainty quantification
%, presents another promising avenue for research. 
%Finally, we intend to conduct more extensive case studies across a broader range of simulation domains to further validate and refine FabGuard's effectiveness in diverse modeling contexts.

%\textbf{Data Availability Statement:} The data that support the findings of this study are openly available in zenodo at \url{https://doi.org/10.5281/zenodo.13360463} 

%\textbf{Conflict of interests statement:} Authors have  no relevant financial or non-financial competing interests to report.

\bibliographystyle{alpha}
\bibliography{miv}

\end{document}